\documentclass[osajnl,twocolumn,showpacs,superscriptaddress,10pt]{revtex4-1}
\usepackage{amsmath,amssymb,graphicx}
\usepackage[center]{subfigure}

\begin{document}

\title{Direct femtosecond pulse compression with miniature-sized Bragg cholesteric liquid crystal}
\author{Liyan Song}
\address{State Key Laboratory of Optoelectronic Materials and Technologies,
Sun Yat-sen University, Guangzhou 510275, China}
\author{Shenhe Fu,$^{1}$ Yikun Liu,$^{1,*}$ Jianying Zhou,$^{1,\dag}$ Vladimir G. Chigrinov}
\address{Hong Kong University of Science and Technology, Clear Water Bay, Kowloon, Hong Kong}
\author{Iam Choon Khoo}
\address{Electrical Engineering Department, Pennsylvania State University, University Park, Pennsylvania, USA\\
$^{*}$Corresponding author: ykliu714@gmail.com \\
$^{\dag}$Corresponding author: stszjy@mail.sysu.ude.cn}
\begin{abstract}
Direct compression of femtosecond optical pulses from a Ti:sapphire laser oscillator was realized with a cholesteric liquid crystal acting as a nonlinear 1D periodic Bragg grating. With a 6-$\mu$m thick sample, the pulse duration could be compressed from 100 to 48 femtoseconds. Coupled-mode equations for forward and backward waves were employed to simulate the dynamics therein and good agreement between theory and experiment was obtained.\newline
\end{abstract}

\maketitle

\noindent Femtosecond optical pulses are widely used for scientific research and industrial applications. The laser oscillator usually delivers an output with a fixed pulse duration, and shortening of the pulse duration is in general realized with a pulse compression scheme. Although effective pulse compression can be achieved with grating pair \cite{1}, prism-pair \cite{2} in conjunction with specific optical fibers \cite{3,4,5,6}, the need to integrate multiple optics elements degrades the system robustness \cite{7,8}. Another strategy is to employ a self-compression scheme, best exemplified by soliton excitation in Bragg grating structures \cite{9,10,11,12} where a compression rate as high as 5 was realized \cite{5}. However, the works reported so far are focused on the compression of pulses of several to tens of picosecond \cite{9,10,11,12} durations. In order to achieve a sizeable compression, the strength of material dispersion and nonlinearity should be well matched. For fiber Bragg gratings (FBGs), typical nonlinear coefficient is in the order of 10$^{-16}$ cm$^{2}$/W; in conjunction with a peak power of kilowatts \cite{11}, the resulting nonlinear effective length is in the millimeter scale \cite{12}. Therefore in order to match the comparable dispersion effect, usually picosecond pulses were adopted \cite{10,11}. Recently, a large nonlinear coefficient in the order of 10$^{-12}$ cm$^{2}$/W with femtosecond (fs) response time was reported in a cholesteric liquid crystals (CLC) system \cite{13,14}, which essentially acts as a 1-D Bragg grating \cite{15}. This nonlinearity could markedly reduce the effective nonlinear length to micrometer \cite{10,16,17}, hence allowing the compression of pulses in the order of tens to hundreds of femtosecond.\newline
\indent In this Letter, direct compression of femtosecond pulses was observed for the first time in a compact CLC cell with peak intensity in the order of 1 GW/cm$^{2}$. The large nonlinearity \cite{13,18} in this Bragg grating could induce sufficient phase modulation at high intensity to compensate for the group velocity dispersion (GVD) and compress the femtosecond pulse. In our experiment, a glass-substrate cell with a gap size of 6  $\mu$m is used, and the thin cell length was chosen to match the longitudinal spatial extent of the femtosecond laser. The left-handed helical structure in planar texture was obtained from a mixture of nematics (5CB, Slichem) with chiral agent (R1011, Slichem). With the 7.5-wt\% concentration of chiral dopant, the transmission spectrum of the CLC cell exhibits a stop band centered around 778 nm (at room temperature (Fig. 1(a)). By adopting the average refractive index and modulation depth $\Delta n$ = 0.19 of 5CB \cite{14}, a pitch around 475 nm and bandwidth of 90 nm was obtained based on Bragg reflection [19]. This bandwidth is much broader than those of FBGs \cite{11,12}, and sufficient GVD can be obtained near the edges of the stop-band for fs pulse compression purposes.
\begin{figure}[tbh]
\centering
\includegraphics[width=8.4cm, height=3.5cm]{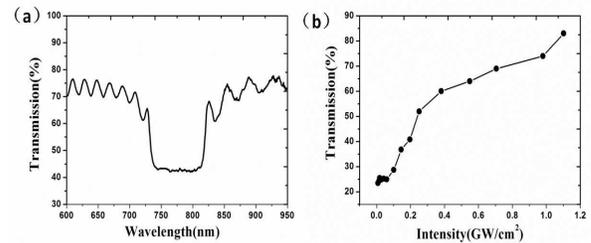} %
\caption{(Color online) (a) Transmission spectrum of CLC from linearly polarized light; (b) Transmission variation with the input laser intensity with the wavelength located at 815 nm. The 815 nm is left-handed circularly polarized}
\label{fig1}
\end{figure}\newline
\indent Our measurements show an increasing intensity dependence for the transmittance of light at the long wavelength edge of CLC (Fig. 1(b)), indicating a blue-shift of band structure caused by the nonlinear index modulation \cite{13} at high laser intensity. This is corroborated by a Z-scan \cite{14,20,21} performed at the same wavelength. The nonlinear coefficient was measured to be about $-$10$^{-11}$ cm$^{2}$/W [note the negative sign]. This is in line with Ref \cite{14} and is at least four orders of magnitude higher than a fiber Bragg grating \cite{9,10,11,12}. Such large nonlinearity and the highly localized density of mode (DOM) near the band edge of PBG material \cite{13,21} enable the observed femtosecond pulse compression effect.\newline
\indent In our experiment, a Ti:sapphire femtosecond mode-locked laser (Tsunami, Spectra-Physics) with average laser output power of 3 W, a repetition rate of 80 MHz and pulse width of 100 fs was employed as the pulse source. For a 6 $\mu$m CLC cell in the experiment, a nonlinear length $L_{NL}=$ 12 $\mu$m \cite{16,17} was obtained. Using the GVD of typical Bragg grating \cite{11}, we estimated a dispersion length $L_{D}$=10 $\mu$m \cite{17}, which matches well with the nonlinear strength. In the pulse compression measurement, a left-handed circularly polarized light at 815 nm located at long-wavelength edge of the stop band was applied as the input. Its pulse duration was measured from the autocorrelation traces using a non-collinear second harmonic autocorrelation scheme with a BBO crystal (a Glan-Taylor prism was used to select a linear polarization at the exit of the CLC cell), as shown in Fig. 2.
\begin{figure}[tbh]
\centering
\includegraphics[width=8.4cm, height=3.5cm]{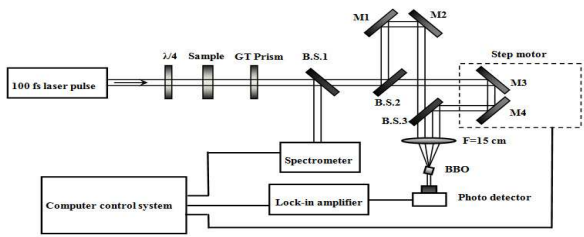} %
\caption{(Color online) Schematic of experimental setup for femtosecond pulse compression with a cholesteric liquid crystal. G.T.Prism, Glan-Taylor Prism; B.S., beam splitter, M, mirror; BBO, Barium Boron Oxide.}
\label{fig1}
\end{figure}\newline
\indent The focused Gaussian beam radius (at $1/e^{2}$ point) was around 0.1 mm and typical intensity was in the order of 1 GW/cm$^{2}$. Fig. 3 demonstrates the femtosecond pulse compression after propagating in the 6 $\mu$m CLC cell, with pulse duration (full width at half maximum, FWHM) of 48 fs at 1.04 GW/cm$^{2}$, i.e., a compression ratio greater than 2 is obtained.\newline
\indent The process can be qualitatively described as the interaction between GVD and induced nonlinear phase modulation. In the absence of significant nonlinearity at low input intensity [19], the femtosecond laser pulse would experience broadening due to the Bragg grating dispersion. At a high input laser intensity of 1.04 GW/cm$^{2}$ the induced phase modulation will be sufficient to compensate for the GVD. For a more detailed theoretical analysis, we adopted a coupled-mode theory model \cite{17} for the dynamic evolution of ultrashort pulses in a periodic structure to simulate the nonlinear CLC system.
\begin{figure}[tbh]
\centering
\includegraphics[width=7cm, height=5cm]{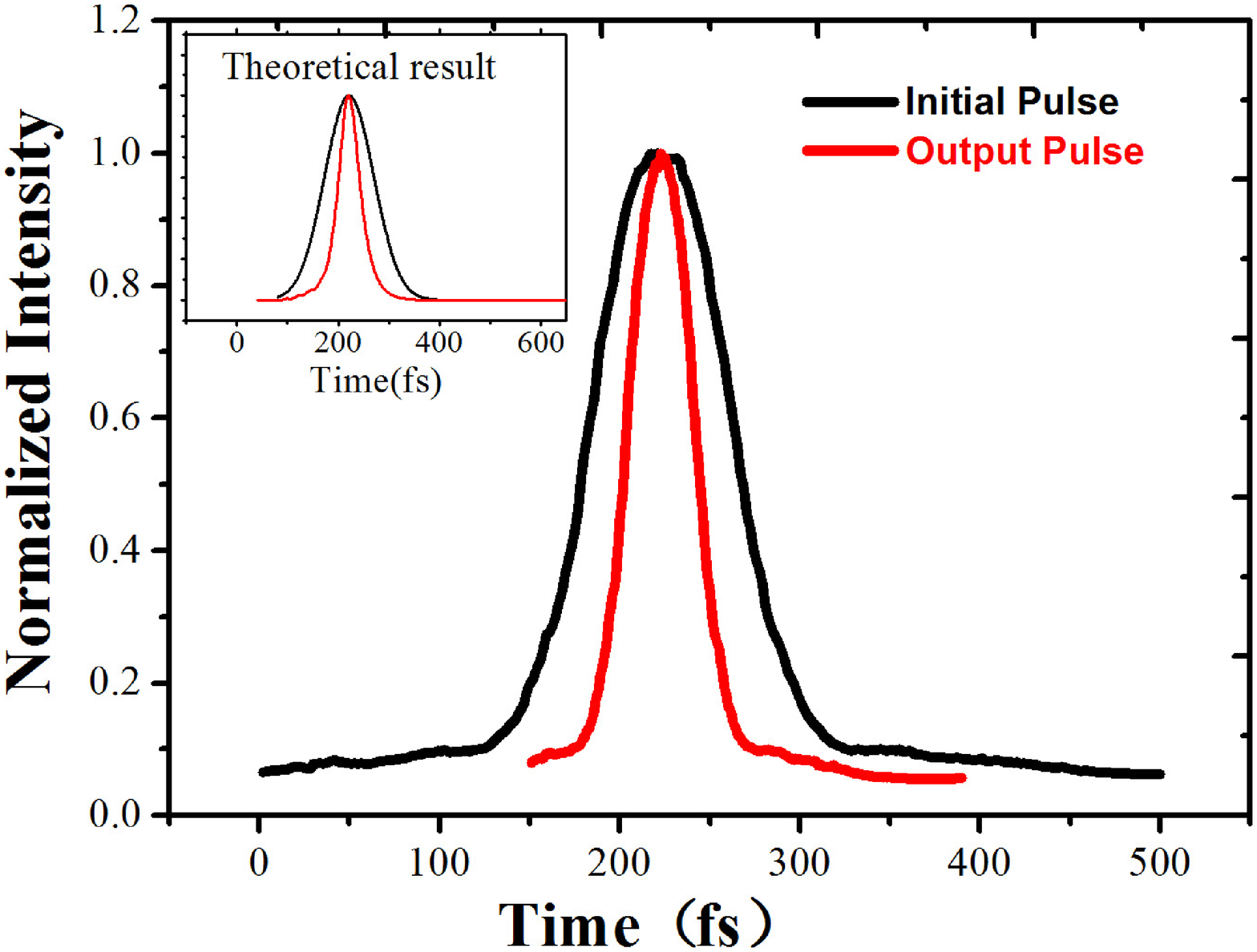} %
\caption{(Color online) Experimental results of femtosecond pulse compression. Initial transform limited pulse (100 fs) is shown as a black line, while the output profile corresponding to an input intensity of 1.04 GW/cm${^2}$ is shown in red line (48 fs); the inset figure corresponds to the simulation results using the measured experimental parameters.}
\label{fig1}
\end{figure}\newline
\indent In the case of a left-handed circularly polarized light, the refractive index in the helical Bragg grating structure could be expressed as
\begin{equation}
n=n_{eff}+\Delta n\cos(2\pi z/\Lambda)+n_{2}|E^{CP}|^2,
\end{equation}
where $n_{eff}$ denotes the average refractive index of CLCs without modulation, $\Delta n$ is depth of linear refractive index modulation, $\Lambda$ is the period, which is one half of the pitch of CLCs, $n_{2}$ is the nonlinear Kerr coefficient while the superscript $CP$ of electric field intensity represents circularly polarization. For the slowly varying envelopes of forward and backward waves, the coupled-mode equations become
\begin{gather}
\frac{1}{v_{g}}\frac{\partial E_{\pm}^{CP}}{\partial t}=\mp\frac{\partial E_{\pm}^{CP}}{\partial z}+i\delta E_{\pm}^{CP}+i\kappa E_{\mp}^{CP} \notag \\
-i\gamma(|E_{\pm}^{CP}|^{2}+2|E_{\mp}^{CP}|^{2})E_{\pm}^{CP},
\end{gather}
where $E_{+}^{CP}$ and $E_{-}^{CP}$ denote the forward and backward waves, respectively. $v_{g}=c/n_{eff}$ is the pulse group velocity in the absence of nonlinearity, with $c$ being the light's velocity in vacuum; $\delta$ is wave-number detuning and $i$ is the imagivative number; $\kappa=\pi\Delta n/\lambda$ is the coupling coefficient between forward and backward waves, with $\lambda$ being the injected wavelength; $\gamma=kn_{2}=2\pi n_{2}/\lambda$ is the nonlinear parameter related to Kerr coefficient $n_{2}$. In our numerical simulation, $n_{2}$ was set at $-1.5\times$10$^{-11}$cm$^{2}$/W, as measured in this experiment. The coupled-mode equations were solved by the Split-step Fourier method \cite{17}, using experimental parameters as input. The numerical result of pulse compression was displayed in the inset in Fig. 3, showing good agreement with experimental observations.
\begin{figure}[tbh]
\centering
\includegraphics[width=7cm, height=5cm]{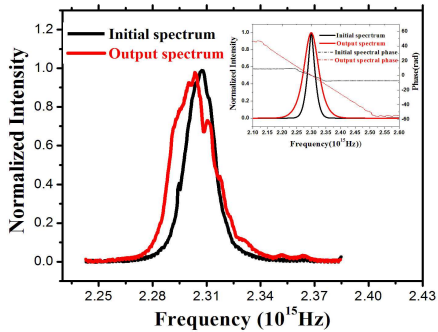} %
\caption{(Color online) Experimental results of spectrum broadening. Initial pulse spectrum is shown as a black line, while that of output pulse with injected intensity of 1.04 GW/cm$^{2}$ as a red line; the inset figure corresponds to simulation results of spectrum (solid lines) and spectral phase (dashed lines).}
\label{fig1}
\end{figure}\newline
\indent To further verify the pulse compression in time domain, the spectrum was analyzed using a grating spectrometer with a resolution around 0.1 nm in the measured spectral region. We found that both the shape and width of spectrum was well maintained at low input intensity with no evidence of nonlinear spectral broadening. On the other hand, a broadened spectrum with new frequencies components were generated at higher input intensity, as depicted in Fig. 4. The FWHM of the spectrum increased from 15.7 THz of source pulse to 29.1 THz for the compressed pulse at an input intensity of 1.04 GW/cm$^{2}$. This spectral broadening is consistent with the temporal compression of the laser pulse. The simulated results of spectrum broadening in the inset in Fig. 4 showed a good agreement in term of shapes both in time and spectral domain. A further calculation shows that the time-bandwidth product in two situations varies within 10\%, close to the transform limit of the femtosecond pulses. This is again verified by the linear spectral phase of initial and output pulses shown in dashed lines in the inset of Fig. 4. \newline
\indent We have also investigated the intensity dependence of the pulse compression. Fig. 5 shows the experimental results together with the simulation plot. Pulse compression begins to be measurable for input intensity above 0.825 GW/cm$^{2}$. The compression rate increases with higher input intensity until the onset of pulse splitting, which places a limit on the achievable compression rate \cite{10,11,12}. This limiting effect was indeed observed in our experiments; as a function of laser intensity, the pulse duration continued to decrease and finally tapered off at a compression rate of 2. These experimental observations are in good agreement with our theoretical simulation, which indicates that our sample nonlinearity and thickness would enable a compression rate of 2.3 before pulse splitting. The slight deviation between theory and experiment for the achievable compression rate is attributed to the energy loss induced from interaction in the gap between CLC and the ITO substrate (cell window) at high laser intensity; another likely factor is imperfection in the CLC structure.
\begin{figure}[tbh]
\centering
\includegraphics[width=7cm, height=5cm]{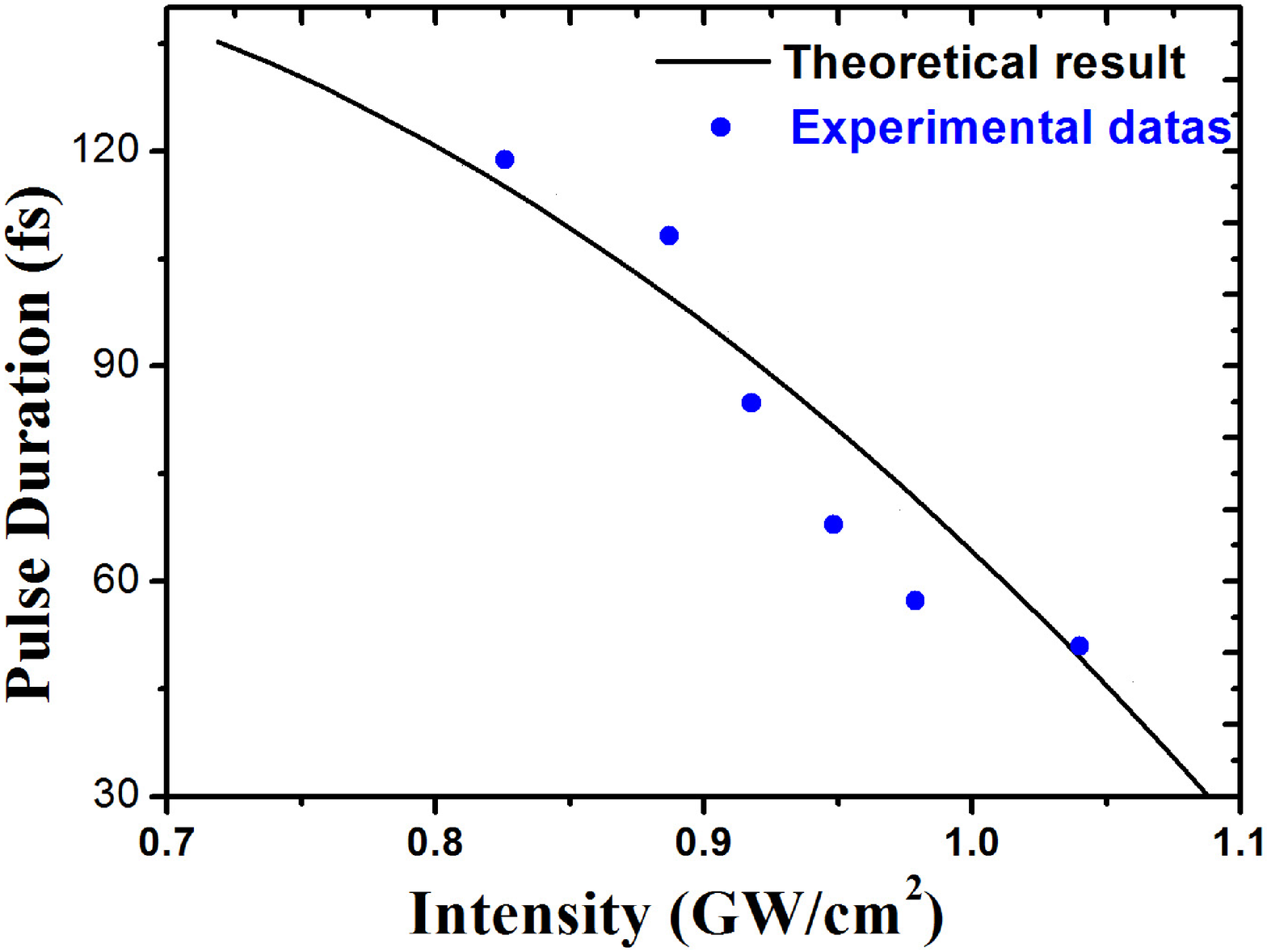} %
\caption{(Color online) Experimental and theoretical results for the pulse compression.}
\label{fig1}
\end{figure}\newline
\indent In conclusion, we have demonstrated for the first time to our knowledge sizeable compressions of femtosecond laser pulses using a 6 $\mu$m thick cholesteric liquid crystal cell. As such thin optical element can be easily integrated into an optical path, this low-threshold and efficient compact device is expected to have potential applications in high-speed optical information processing.

\section{Acknowledgement}
\indent The authors acknowledge the financial support by the National Key Basic Research Special Foundation (G2010CB923204), Chinese National Natural Science Foundation (10934011). I. C. Khoo's work is supported by the US Air Force Office of Scientific Research.

\end{document}